\documentclass[preprint,showkeys,secnumarabic,amsfonts,showpacs,amsmath,amssymb]{revtex4}
\usepackage[dvips]{color}
\usepackage{epsfig}
\usepackage{amsmath}
\usepackage{graphicx}
\def\Box{\hbox{$\rlap{$\sqcup$}\sqcap$}}

\begin{document}
\title{ Investigation of Bouncing Universe and Phantom Crossing in Modified Gravity Coupled with Weyl Tensor and its Reconstruction}

\author{M. Ghanaatian}\email{m_ghanaatian@pnu.ac.ir}\affiliation{Department of Physics, Payame Noor University, Iran}
\author{F. Milani}\email{fmilani@guilan.ac.ir}\affiliation{Department of Physics, Payame Noor University, Iran}

\date{\today}

\begin{abstract}
\noindent \hspace{0.35cm}

In this study, FRW cosmology in modified gravity containing
arbitrary function $f(R)$ is taken into consideration when our
action are coupled with Weyl tensor. It is indicated that the
bouncing solution emerges in the model while the equation of state
(EoS) parameter crosses the phantom divider. In this research,
cosmological usage of the most promising candidates of dark energy
in the framework of $f(R)$ theory coupled by Weyl tensor is
explored. A $f(R)$ gravity model in the spatially flat FRW
universe acoording to the ordinary version of the holographic dark
energy model, which describes accelerated expansion of the
universe is reconstructed. The equation of state parameter of the
corresponding Weyl gravity models are obtained as well. We
conclude that the holographic and Weyl gravity models can behave
like phantom or quintessence models, whereas the equation of state
parameter of the models can transit from quintessence state to
phantom regime as shown recent observations.
\end{abstract}

\pacs{04.50.Kd; 98.80.-k}

\keywords{Modified gravity; Bouncing universe; $\omega$ crossing;
Weyl gravity; Reconstructing}

\maketitle

\section{Introduction}
One explanation for physicists with respect to how our universe
formed is the big bang theory, commencing with Albert Einstein's
theory of general relativity and bouncing theory in an alternative
way of looking at how the universe commenced. The big bang theory
tries to enunciate what the universe looked like before the
planets and the stars came to existence. The universe, about 13.7
billion years ago, is believed by scientists to have condensed
into a small region of matter and energy called a singularity. All
of a sudden, the singularity exploded, expanding at an incredibly
fast rate. Astronomers are of the belief that following the big
bang, the universe was like a soup with small slices of matter
floating around. The matter combined to create protogalaxies,
which in turn combined to form galaxies. Within those galaxies,
gas and dust blended to make stars. And around those stars,
gravity pulled pieces of matter together to form planets.

The theory is simply the one which can not be recreated or proved
by scientists at this point in time. Moreover, some detractors
refer to a few weaknesses in big bang thinking. This signifies
that other theories have been suggested for its replacement.
Cosmologists are still attempting to predict the fate of our
universe, trying to come to this conclusion whether it will expand
forever, stabilize or collapse in on itself. Some cosmologists
believe that the universe will eventually grow no more. Our
universe will collapse in on itself into a singularity as gravity
pulls matter down, an event billions of years from now which is
called the big crunch. The planets, stars and galaxies in and of
themselves are not dense enough to cause the big crunch. However,
cosmologists believe that unseen materials exists and may exert
enough gravitational force to stop the universe's expansion and
cause the big crunch. The bouncing theory combines the big bang
and big crunch theories to develop a vision of an infinite,
cyclical cosmos in which the universe over and again expands from
a singularity only to ultimately collapse back in on itself,
before doing it all over again. To put it another way, a bouncing
universe would continuously expand and contract.

It should be mentioned that there is no evidence as to what will
occur to our universe in the future, but about its beginning, it's
differ comparatively. Observational data of type Ia Super-Novae
(SNIa) {\cite{c14-02}} have determined basic cosmological
parameters in high-precisions. They are indicative of the fact
that, the universe is spatially flat and dominated by two dark
components containing dark energy and dark matter, and comprises
nearly $73\%$ dark energy (DE) and $27\%$ dust matter (cold dark
matters plus baryons) with negligible radiations. Simultaneously,
as to the origin of DE, they posed a fundamental problem. The
combined analysis of SNIa {\cite{a1-02}}, contingent upon the
background expansion history of the universe around the redshift
$z<1$ as galaxy clusters measurements and Wilkinson Microwave
Anisotropy Probe (WMAP) data {\cite{c15}}, Sloan Digital Sky
Survey (SDSS) {\cite{c16-02}}, Chandra X-ray Observatory (CXO)
{\cite{c17}} etc., it shows some cross- checked information of our
universe, providing surprising proof as to the fact that the
expansion of the universe for the time begin seems to have
accelerating, behavior, being imputed to dark energy (DE), a
strange energy with negative pressure. In contrast, dark matter
(DM), a matter without pressure, is basically utilized to describe
galactic curves and large-scale structure formation{\cite{c3-02}}.

It is shown by the cosmological acceleration the present day
universe is dominated by smoothly distributed slowly varying DE
component. The constraint derived from SNIa has a degeneracy in
the equation of state (EoS) of DE {\cite{a2-03}}. However, the
nature of DE until now continues to be unknown, people have
suggested some candidates for its explanation. The cosmological
constant, $\Lambda$, in a model in which the universe's equation
has a cosmological constant, indicated by $\Lambda$, and Cold Dark
Matter ($\Lambda$CDM), is the most notable theoretical candidate
of DE, having an equation of state with $\omega = -1$. This
degeneracy is offered even by adding other constraints coming from
Cosmic Microwave Background (CMB) {\cite{a3-02}} and Baryon
Acoustic Oscillations (BAO) {\cite{a4}}. Astronomical observations
denote that the cosmological constant, in their orders of
magnitude, to be much smaller than it is calculated in modern
theories of elementary particles {\cite{karami2}}. Two of the most
notable difficulties faced with the cosmological constant are the
"fine-tuning" and the "cosmic coincidence" {\cite{karami3}}. The
constraints, nowadays, on the EoS around the cosmological constant
value, $\omega= -1 \pm 0.1$ {\cite{c3-02}}-{\cite{c4-02}} and this
probability exists that $\omega$ may differ in time. From the
theoretical point of view there are three essentially different
cases: $\omega>-1$ (quintessence), $\omega = -1$ (cosmological
constant) and $\omega <-1$ (phantom)
({\cite{c5-06}}-{\cite{c8-03}} and refs. therein).

The models of DE can be generally categorized into two groups
{\cite{a5,a6}}. In the first group, a specific matter leading to
an accelerated expansion is introduced. Most of scalar field
models such as quintessence {\cite{a7-05}} and k-essence
{\cite{a8-02}} belong to this class. The second class, considered
in this study, corresponds to the so-called modified gravity
models such as $f(R)$ gravity {\cite{a9-03}}, scalar-tensor
theories {\cite{a10-11}} and brane-world models{\cite{a11-02}}. In
order to break the degeneracy of observational constraints on
$\omega$ and to discriminate between a DE models, it is important
to find additional information other than the background expansion
history of the Universe {\cite{a12}}.

Modified gravity, in the second classification {\cite{b1}},
suggests fine alternative for DE origin. The expectation is that
gravitational action has some extra terms which became relevant
recently with the significant decrease of the universe curvature.
The modified gravity can be obtained in three ways: first by
substituting scalar curvature $R$, or by $f(R)$, second by taking
additional curvature invariant terms into account like
Gauss-Bonnet (GB) term as $\mathcal{G}$ or $f(\mathcal{G})$
{\cite{prd}}, and third by replacing a coupling of two methods ago
as $f(R,\mathcal{G})$, in the Einstein-Hilbert action
($I_{EH}=\frac{1}{16\pi G}\int{d^4 x\sqrt{-g}R}$) where $G$ is
gravitational constant.

In this regard, for a long time, conformal transformations and
conformal techniques have been broadly in use in general
relativity (Ref. {\cite{Takook1}} and references therein). It has
often been claimed that conformal invariant field theories are
renormalizable {\cite{Takook2}} and conformal gravity may be an
alternative theory of gravity {\cite{Takook3}}. Because the
gravitational field is long range and appears to travel with the
speed of light, in the linear approximation, at least, it is
expected for the equations to be conformally invariant. Einstein's
theory of gravitation, known by all, is not conformally invariant.
This theory appears not to be a totalizing universal theory of
gravitational field because of the mentioned fact and some other
issues emanating from standard cosmology and quantum field theory
{\cite{Takook4}}. Many have tried to generalize this theory which
goes back to the early days of general relativity (for reviewing
see {\cite{Takook5}}). The first invariant gravitational theory
under the scale transformation was presented by Weyl, being called
Weyl gravity. The Conformal Weyl gravity is reliant upon local
conformal invariance of the metric of the form
$g_{\mu\nu}(x)\rightarrow \Omega^2(x)g_{\mu\nu}(x)$, being
conducive to a theory with the field equation of fourth order
derivative (higher-derivative theories) where $\Omega$ is
independent of the space-time coordinates.

Some metric formulation of modified $f(R)$ gravities are suggested
{\cite{b1,prd}}, {\cite{b2}}-{\cite{b5-12}} describing the origin
of cosmic acceleration. Particular attention is paid to $f(R)$
models {\cite{b6-03}}-{\cite{b9}} with the effective cosmological
constant phase because such theories may easily reproduce the
well-known $\Lambda\text{CDM}$ cosmology. Such models subclass
which does not violate Solar System tests represents the real
alternative for standard General Relativity{\cite{b10}}.

The Friedman equation, on the other hand, constitutes the starting
point for nearly all researches in cosmology. The Friedman
equation has been, corrected during the past few years being
proposed in varying contexts, generally inspired by brane-world
investigation {\cite{c9, c10}}. These changes are often of a type
that involves the total energy density $\rho$. In {\cite{c11}},
multi-scalar coupled to gravity is studied in the context of
conventional Friedman cosmology. It is found that the cosmological
trajectories can be viewed as geodesic motion in an increased
target space.

There are several phenomenological models which describe the
crossing of the cosmological constant barrier {\cite{c12-05,
c13-04}}. Finding a model following from the basic principles is
of importance and which describes a crossing of the $\omega = -1$
barrier.

In this paper, in section 2, the dynamics of the FRW cosmology in
modified gravity is considered. We discuss analytically and
numerically a detailed examination of the conditions for having
$\omega$ across over $-1$. The necessary conditions required for a
successful bounce is discussed in this section as well. In section
3 we will reconstruct our model corresponding to the Holographic
Dark Energy (HDE) respectively. Finally, we summaries our paper in
section 4.

\section{The Model}

In the $f(R)$ theory of gravity the Einstein-Hilbert action is
replaced by the square of the conformal Weyl tensor
\begin{eqnarray}\label{ac1}
I_{W}=-\frac{\alpha}{4}\int{d^4x\sqrt{-g}\left\{C_{\mu\nu\rho\lambda}C^{\mu\nu\rho\lambda}+2f(R)\right\}},
\end{eqnarray}
where $\alpha=1/8\pi G$ and $C_{\mu\nu\rho\lambda}$ is the Weyl
tensor
\begin{eqnarray}\label{Weyl tensor}
C_{\mu\nu\rho\lambda}=R_{\mu\nu\lambda\rho}
-\frac{1}{2}(g_{\mu\lambda}R_{\nu\rho}-g_{\mu\rho}R_{\nu\lambda}-g_{\nu\lambda}R_{\mu\rho}+g_{\nu\rho}R_{\mu\lambda})
+\frac{R}{6}(g_{\mu\lambda}g_{\nu\rho}-g_{\mu\rho}g_{\nu\lambda})\cdot
\end{eqnarray}
The action (\ref{ac1}) can be written as flows
\begin{eqnarray}\label{ac2}
I_{W}&=&-\frac{\alpha}{4}\int{d^{4}x\sqrt{-g}\left\{R^{\mu\nu\rho\lambda}R_{\mu\nu\rho\lambda}-2R^{\mu\nu}R_{\mu\nu}+\frac{1}{3}R^{2}
+2f(R)\right\}},
\end{eqnarray}
since
$\sqrt{-g}(R^{\mu\nu\rho\lambda}R_{\mu\nu\rho\lambda}-4R^{\mu\nu}R_{\mu\nu}+R^{2})$
is a total divergence (Gauss-Bonnet term), it does not contribute
to the equation of motion and one can simplify the action as
follows
\begin{eqnarray}\label{ac3}
I_{W}&=&-\frac{\alpha}{2}\int{d^{4}x\sqrt{-g}\left\{R_{\mu\nu}R^{\mu\nu}-\frac{1}{3}R^{2}+f(R)\right\}}\nonumber\\
&\equiv&-\frac{\alpha}{2}\int{d^{4}x\left\{\mathcal{W}^{(2)}-\frac{1}{3}\mathcal{W}^{(1)}+\sqrt{-g}f(R)\right\}}\cdot
\end{eqnarray}

The total action is $I \equiv I_W + I_M$, where $I_M$ is the
conformal matter action. Functional variation of the total action
with regard to the matter fields produces the equations of motion
while its functional variation considering the metric generates
the $f(R)$ modified gravity coupled by Weyl field equation.
Therefore, taking the variation of the action (\ref{ac3}) with
respect to the metric $g^{\mu\nu}$ , the field equations can be
obtained as {\cite{Nojiri-1},\cite{Starobinsky}}
\begin{eqnarray}\label{G}
R_{\mu\nu}-\frac{1}{2}R
g_{\mu\nu}=\frac{1}{\alpha}T_{\mu\nu}^{(R)}\cdot
\end{eqnarray}
where
\begin{eqnarray}\label{Tmunu}
\frac{1}{\alpha}T_{\mu\nu}^{(R)}=\frac{1}{2}g_{\mu\nu}f(R)-R_{\mu\nu}f'(R)+(\nabla_{\mu}\nabla_{\nu}-g_{\mu\nu}\Box)f''(R)+\mathcal{W}_{\mu\nu},
\end{eqnarray}
and
\begin{eqnarray}
\mathcal{W}_{\mu\nu}&\equiv&\mathcal{W}^{(2)}_{\mu\nu}-\frac{1}{3}\mathcal{W}^{(1)}_{\mu\nu}\nonumber\\
&=&-\frac{1}{2}g_{\mu\nu}\Box R-\Box
R_{\mu\nu}+\nabla_{\rho}\nabla_{\mu}R_{\nu}^{\rho}+\nabla_{\rho}\nabla_{\nu}R_{\mu}^{\rho}-2R^{\rho}_{\mu}R_{\nu\rho}
+\frac{1}{2}g_{\mu\nu}R_{\rho\lambda}R^{\rho\lambda}\nonumber\\
&-&\frac{1}{3}(2\nabla_{\mu}\nabla_{\nu}R-2g_{\mu\nu}\Box
R-2RR_{\mu\nu}+\frac{1}{2}g_{\mu\nu}R^2)\cdot
\end{eqnarray}

Here $R_{\mu\nu}$ is the Ricci tensor, respectively. Also the
prime is also indicative of a derivative with respect to R. Now if
we consider the spatially flat FRW metric for the universe as
\begin{eqnarray}
ds^2=-dt^2+a^2(t)\sum _{i=1}^{3}(dx^i)^2,
\end{eqnarray}
and $ T_{\mu\nu}^{(R)}=g_{\mu\nu}T_{\mu}^{\nu (R)}$ then the set
of field equations (\ref{G}) reduce to the modified Friedmann
equations in the framework of $f(R)$-gravity as
\begin{eqnarray}
3H^2 &=& \frac{\rho_{R}}{\alpha},\label{f1}\\
-2\dot{H}-3H^2&=&\frac{p_{R}}{\alpha}\cdot\label{f2}
\end{eqnarray}

The model can be considered as a standard model with the effect of
the Weyl and $f(R)$ gravity modification contributed in the energy
density and pressure of the Friedman equations. After some
algebraic calculation, the field Eq. (\ref{Tmunu}), when
\begin{eqnarray}\label{ricci scalar}
R=6\dot{H}+12H^2,
\end{eqnarray}
corresponding to standard spatially-flat FRW universe for the $00$
and $ii$ components yields,
 \begin{eqnarray}
\frac{\rho_{R}}{\alpha}&=&-\frac{1}{2}f(R)+3(\dot{H}+H^2)f'(R)-3H\dot{R}f''(R)+\mathcal{W}_{00},\label{rho}\\
\frac{p_{R}}{\alpha}&=&\frac{1}{2}f(R)-(\dot{H}+3H^2)f'(R)+(\ddot{R}+2H\dot{R})f''(R)+\dot{R}^2
f'''(R)+\frac{\mathcal{W}_{ii}}{a^2(t)},\label{p}
\end{eqnarray}
where
\begin{eqnarray}
\mathcal{W}_{00}&=&3\left(\dddot{H}(1-H)-4\ddot{H}H^2+2\dot{H}^2(1-2H-4H^2)-8H^4\right),\label{W00}\\
\frac{\mathcal{W}_{ii}}{a^2(t)}&=&4\ddot{H}(6H^2+H+3\dot{H})+\dot{H}\left(19H^2-12H
+\frac{3}{2}\dot{H}+\frac{9}{2}\right)+\frac{3}{2}\left(3+H^2\right)\cdot\label{Wii}
\end{eqnarray}
Here $H=\frac{\dot{a}}{a}$ is the Hubble parameter and the dot
denotes a derivative with respect to cosmic time $t$. Also
$\rho_{R}$ and $p_{R}$ are the curvature contribution to the
energy density and pressure.

The energy conservation laws are still given by
\begin{eqnarray}
\dot{\rho}_{R} +3H\rho_{R}(1 +\omega)&=&0,\label{rhoR}
\end{eqnarray}
where $\omega=\frac{p_{R}}{\rho_{R}}$ is the equation of state
(EoS) parameter due to the curvature contribution which defined as
{\cite{sadeghi}} and it's given by
\begin{eqnarray}
\omega=-1-\frac{6f'''\mathcal{A}^2+6f''(\mathcal{B}+2H\mathcal{A})
-f'(\dot{H}+3H^2)+\frac{1}{2}f+4\ddot{H}\mathcal{C}+\dot{H}\mathcal{D}+\frac{9}{2}+\frac{3}{2}H^2}
{18f''\mathcal{A}H-3f'(\dot{H}+H^2)+\frac{1}{2}f+3\mathcal{E}+24H^2-\frac{1}{12}\dot{H}^2},
\end{eqnarray}
where
\begin{eqnarray}
\mathcal{A}&\doteq&\ddot{H}+4H\dot{H},\nonumber\\
\mathcal{B}&\doteq&\dddot{H}+4H\ddot{H}+4\dot{H},\nonumber\\
\mathcal{C}&\doteq&3\dot{H}+6H^2+H,\nonumber\\
\mathcal{D}&\doteq&19H^2-12H+\frac{3}{2}\dot{H}+\frac{9}{2},\nonumber\\
\mathcal{E}&\doteq&-(1-H)\dddot{H}+4H^2(\ddot{H}+2)+8\dot{H}^2(H+\frac{1}{2})^2\cdot
\end{eqnarray}
In the case of $f(R)=0$ , from Eqs. (\ref{rho}) and (\ref{p}) we
have $\rho_{R} = 0$ and $p_{R} = 0$. Therefore Eqs. (\ref{f1}) and
(\ref{f2}) transform to the usual Friedmann equations in GR. But
for an arbitrary $f(R)$ as,
\begin{eqnarray}
f(R)=\frac{1}{2}\partial_{\mu}\phi\partial_{\nu}\phi
g^{\mu\nu}+\frac{1}{12}\phi^2 R,\label{F}
\end{eqnarray}
in a FRW cosmological model, for only time dependent $\phi$ by
invariance of the action under changing fields and vanishing
variations at the boundary, the equations of motion for scalar
fields $\phi$ become,
\begin{eqnarray}
\ddot{\phi}+3H\dot{\phi}-\frac{1}{3}\phi R=0\cdot\label{EOM}
\end{eqnarray}
and the equation of state (EoS) parameter is given by
\begin{eqnarray}
\omega=-1-\frac{(1-H)\dddot{H}+\frac{1}{3}\mathcal{F}\ddot{H}
+\frac{1}{12}\mathcal{G}\dot{H}-\frac{5}{2}(5H^2-1)}{\mathcal{E}-\frac{1}{12}(\ddot{\phi}+\phi^2)},
\end{eqnarray}
where
\begin{eqnarray}
\mathcal{F}&\doteq&4H(1+3H)+12\dot{H},\nonumber\\
\mathcal{G}&\doteq&19H^2-12H-(48H^2+12H)\dot{H}+\frac{1}{6}\phi^2+\frac{9}{2}\cdot
\end{eqnarray}
Also the solution for $H(t)$, Eq. (\ref{f1}), provides a dynamical universe with contraction for $t<0$, bouncing at $t=0$ and then expansion for $t>0$. The above analysis clearly can be seen in the numerical calculation given in Fig. 1.\\
\begin{tabular*}{2.5 cm}{cc}
\includegraphics[scale=.25]{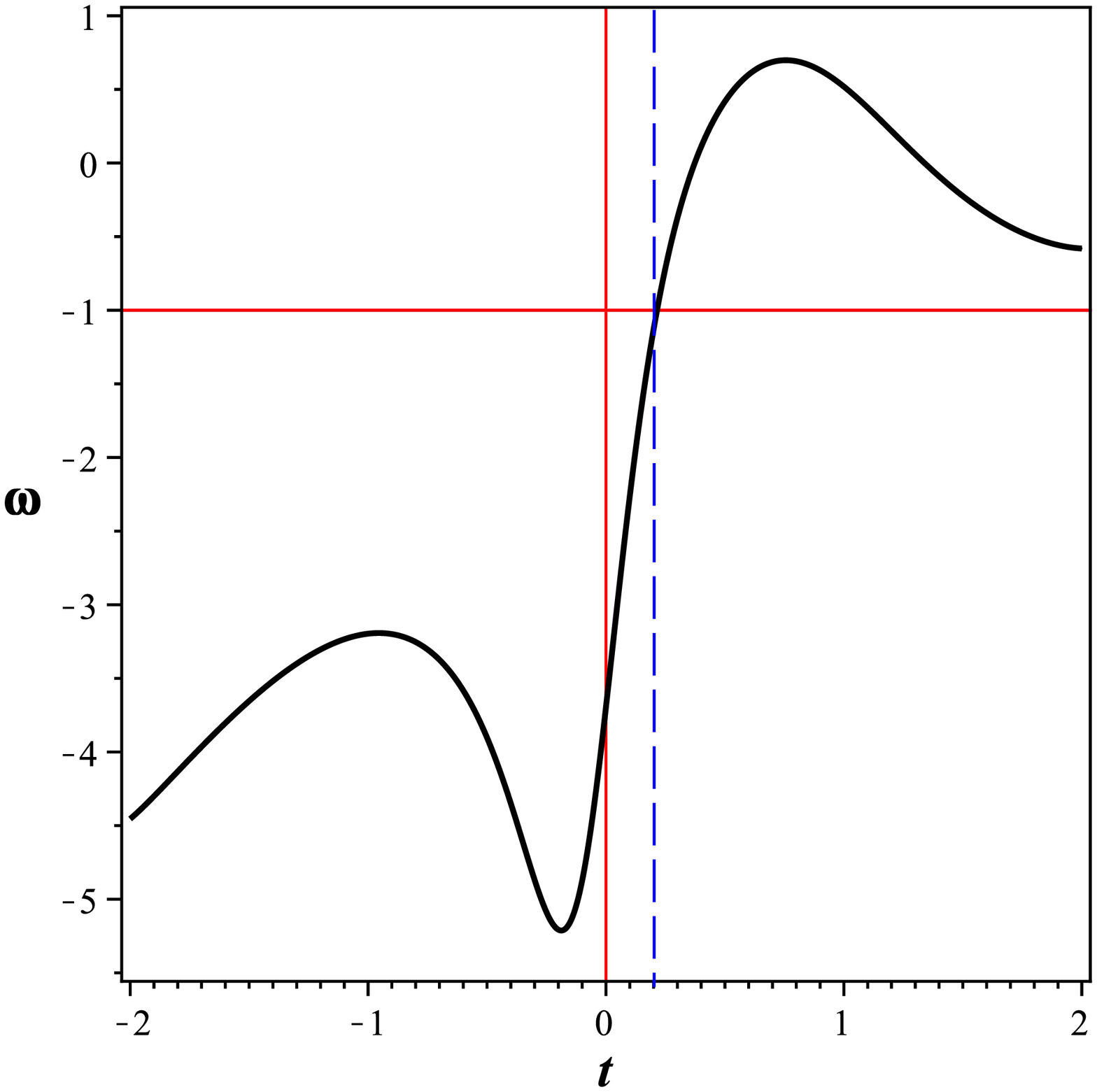}\hspace{0.2 cm}\includegraphics[scale=.25]{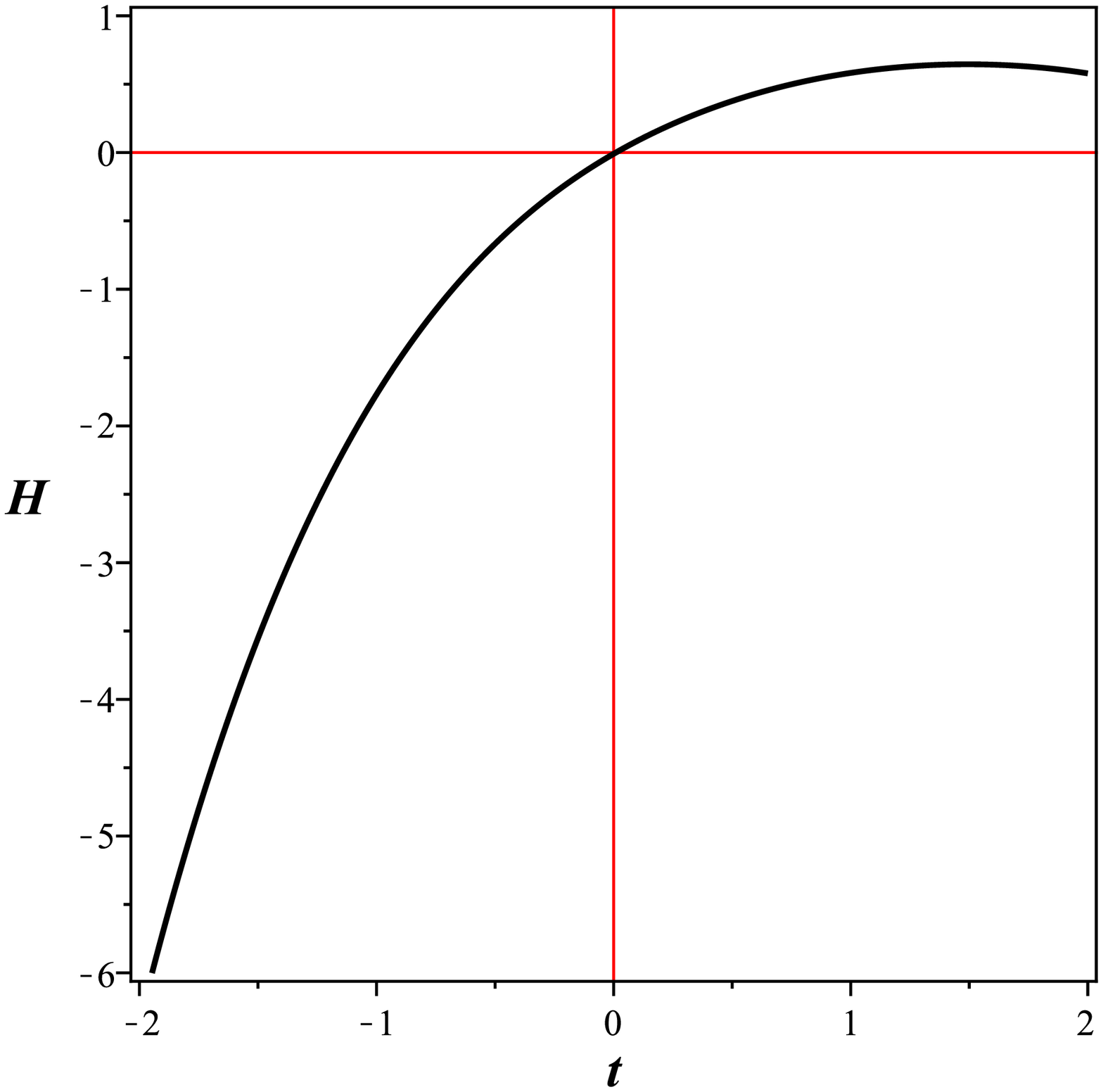}\hspace{0.2 cm}\includegraphics[scale=.25]{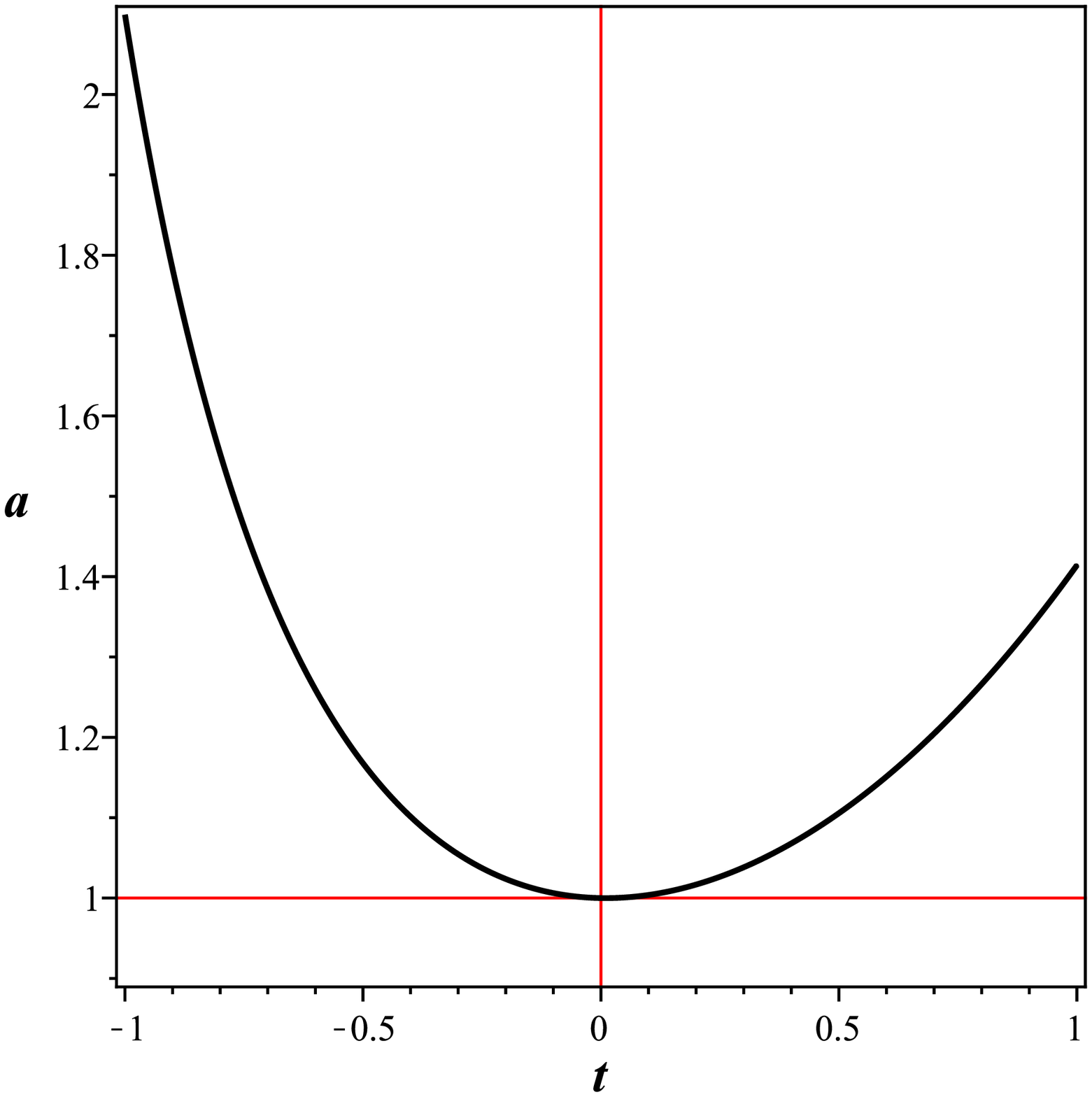}\\
\hspace{1 cm}Fig.1: \, The graph of $\omega$, $H$ and scalar factor $a$,  plotted as function of time.\\
\hspace{1 cm}  Initial values are $\phi(0)=0$ , $\dot{\phi}(0)=0.1$, $\dot{a}(0)=-0.01$ and $\alpha=1$.\\
\end{tabular*}\\

In our model, as {\cite{sadeghi}}, the EoS parameter crosses $-1$
line from $\omega<-1$ to $\omega>-1$, as Fig.1, which is supported
by observations \cite{c8-03}. This model bears the same as quintom
dark energy models which includes two quintessence and phantom
fields \cite{prd}. For a successful bounce implying, a list of
test on the necessary conditions is needed that during the
contracting phase, the scale factor $a(t)$ should being decreased,
i.e., $\dot{a} < 0$, and in the expanding phase, we should have
$\dot{a} > 0$. At the bouncing point, $\dot{a} = 0$, and so around
this point $\ddot{a} > 0 $ for a period of time, the Hubble
parameter $H$ runs across zero from $H < 0$ to $H > 0$ and $H = 0$
at the bouncing point. Around bouncing point, for a successful
bounce, the following condition should be satisfied
\begin{eqnarray}\label{hdot1}
\dot{H}=-\frac{1}{\alpha}(1+\omega)\rho_R>0.
\end{eqnarray}

According to Fig.1, at $t\rightarrow 0$, $\omega<-1$ and $\dot{H}$
is positive and we see that at the bouncing point where the scale
factor $a(t)$ is not zero, we avoid singularity faced in the usual
Einstein cosmology.

At this stage, the cosmological evolution of EoS parameter,
$\omega$, is studied, and we show that, analytically and
numerically, there are conditions that cause the EoS parameter
cross the phantom divide line ($\omega\rightarrow -1$). Let us see
under what conditions the system will be able to cross the barrier
of $\omega=-1$. To do that, one needs $\rho_R + p_R$ to disappear
at a point of ($\phi_0$) and modify the sign after the crossing.
One can accomplish this by requiring  $H (\phi_0) = 0$ and $H$ has
different signs before and after the crossing. For investigating
this possibility, we have to check the condition
$\frac{d}{dt}(\rho_R+p_R)\neq 0$ when $\omega\rightarrow -1$.
Using Eqs. (\ref{F}) in (\ref{rho}) and (\ref{p}) we have,
\begin{eqnarray}\label{rho plus p}
\frac{d}{dt}(\rho_R+p_R)=3(1-H)\ddddot{H}+(\mathcal{F}-3\dot{H})\dddot{H}+\mathcal{J}\ddot{H}-\mathcal{K}\dot{H}^2+\mathcal{M}\neq
0,
\end{eqnarray}
where
\begin{eqnarray}
\mathcal{J}&\doteq&12\ddot{H}+(12H+19)\dot{H}+\mathcal{G},\nonumber\\
\mathcal{K}&\doteq&12\left\{1+(1+4H)\dot{H}\right\}-38H,\nonumber\\
\mathcal{M}&\doteq&\frac{1}{3}\phi\dot{\phi}+3H-96H^3 .
\end{eqnarray}

In this case, our analytical discussion about $\omega\rightarrow
-1$ would be just a boring game on different components of
Eq.(\ref{rho plus p}) to satisfy $\frac{d}{dt}(\rho_R+p_R)\neq 0$.
For example if second and upper orders of derivatives of $H$
respect to cosmic time $t$ would have been vanished, one can find,
\begin{eqnarray}
\dot{H}^2\neq \frac{\phi\dot{\phi}+3H(1-32H^2)}{6(12-19H)}\cdot
\end{eqnarray}

\section{$f(R)$ reconstruction from HDE model}

There are two classes of scale factors which are ordinary
investigated for explaining the accelerating universe in $f(R)$,
$f(\mathcal{G})$ and $f(R,\mathcal{G})$ modified gravities. The
first class of scale factor is donated by \cite{Nojiri, Sadjadi}
\begin{eqnarray}\label{a1}
a(t) = a_0(t_s-t)^{-h},\,\,\,\,\,\,\,\,\,\,\,\,\  t \leq
t_s,\,\,\,\,\,\,\,\,\,\,\,\,\ h > 0\cdot
\end{eqnarray}
and consequently
\begin{eqnarray}\label{H1}
H =\frac{h}{t_s-t}=\sqrt{\frac{h}{6(2h+1)}R}
,\,\,\,\,\,\,\,\,\,\,\,\, \dot{H}=H^2/h ,
\end{eqnarray}
which $\dot{H}> 0$ exhibits the model, matches with a phantom
dominated universe. So, this model is usually so-called the
phantom scale factor in the literature. For the second class
\cite{Nojiri}
\begin{eqnarray}\label{a2}
a(t) = a_0t^{h},\,\,\,\,\,\,\,\,\,\,\,\,\,\,\,\,\,\,\,\,\,\,\,\,\
h > 0,
\end{eqnarray}
with
\begin{eqnarray}\label{H2}
H =\frac{h}{t}=\sqrt{\frac{h}{6(2h-1)}R} ,\,\,\,\,\,\,\,\,\,\,\,\,
\dot{H}=-H^2/h ,
\end{eqnarray}
which $\dot{H}< 0$ shows the model that describes a quintessence
dominated universe. Therefore, this model is so-called the
quintessence scale factor in the literature.

Here we reconstruct the Weyl gravity according to the HDE
scenario. Following {\cite{Li}} the HDE density in a spatially
flat universe is given by,
\begin{eqnarray}\label{rhoLamda}
\rho_{\Lambda}=\frac{12\alpha c^2}{R^2_h},
\end{eqnarray}
where $c=0.818^{+0.113} _{-0.097} $  in recent observational data
used to constrain the HDE model shows the flatten universe
\cite{Li38}. In addition $R_h$ is the future event horizon defined
as
\begin{eqnarray}\label{Rh1}
R_{h}=a\int_{t}^{\infty}{\frac{dt}{a}}=a\int_{a}^{\infty}{\frac{da}{Ha^2}}\cdot
\end{eqnarray}
For the first class of scale factor, Eq.(\ref{a1}), with
Eq.(\ref{H1}), the future event horizon $R_h$ given by
\begin{eqnarray}\label{Rh2}
R_{h}=\frac{1}{h+1}\sqrt{\frac{6h(2h+1)}{R}}\cdot
\end{eqnarray}
Replacing Eq.(\ref{Rh2}) into Eq.(\ref{rhoLamda}) one can get
\begin{eqnarray}\label{rhoLamda1}
\rho_{\Lambda}=\frac{2\alpha c^2(h+1)^2}{h(2h+1)}R \cdot
\end{eqnarray}
Substituting Eq.(\ref{rhoLamda1}) in the differential equation
(\ref{rho}), i.e. $\rho_{R}=\rho_{\Lambda}$, gives the following
solution
\begin{eqnarray}\label{fR1}
f(R)=\lambda_{+}R^{m_+}+\lambda_{-}R^{m_-}+\gamma_{1}R^{\frac{11}{2}}+\gamma_{2}R^{5}+\gamma_{3}R^{\frac{3}{2}}+\gamma_{4}R\cdot
\end{eqnarray}
where
\begin{eqnarray}\label{m}
m_{\pm}=\frac{3+h\pm\sqrt{1-10h+h^2}}{4},
\end{eqnarray}
and
\begin{eqnarray}\label{gamma}
\gamma_{1}&=&8\gamma_c h^2\sqrt{\frac{6h}{2h+1}}\frac{\left(m_+-m_-\right)}{(11-2m_+)(11-2m_-)},\nonumber\\
\gamma_{2}&=&\eta\gamma_c\frac{(m_+-m_-)}{(5-m_+)(5-m_-)},\nonumber\\
\gamma_{3}&=&12\gamma_c (2h+1)\sqrt{\frac{6h}{2h+1}}\frac{\left(m_+-m_-\right)}{(3-2m_+)(3-2m_-)},\nonumber\\
\gamma_{4}&=&-18\gamma_c (2h+1)\frac{(m_+-m_-)}{(1-m_+)(1-m_-)},\nonumber\\
\eta &=&(6 c^2+4)h^4+(15 c^2+4)h^3+(12 c^2-1)h^2+3 c^2 h,\nonumber\\
\gamma_c&=&\frac{-1}{3h^2\sqrt{1-10h+h^2}(2h+1)} \cdot
\end{eqnarray}

Also $\lambda \pm$ are the integration constants that can be
determined from the necessary boundary conditions. Following
{\cite{Nojiri-2}} the accelerating expansion in the present
universe could be generated, if one consider that $f(R)$ could be
a small constant at present universe, that is
\begin{eqnarray}\label{boundary}
f(R_0)=-2R_0 \,\,\,\,\,\,\,\ and \,\,\,\,\,\ f'(R_0)\sim  0\cdot
\end{eqnarray}
where $R_0 \sim (10^{-33}eV)^2$ is the current curvature. Applying
the above boundary conditions to the solution Eq. (\ref{fR1}) one
can obtain
\begin{eqnarray}\label{gammapm}
\lambda_+&=&\frac{2R_0m_--Q(R_0)+m_+P(R_0)}{m_+-R_0m_-},\\
\lambda_-&=&\frac{2R_0m_+-Q(R_0)+m_-P(R_0)}{m_--R_0m_+},
\end{eqnarray}
where
\begin{eqnarray}\label{PQ}
P(R_0)&=&\gamma_{1}R_0^{\frac{11}{2}}+\gamma_{2}R_0^{5}+\gamma_{3}R_0^{\frac{3}{2}}+\gamma_{4}R_0,\\
Q(R_0)&=&\frac{11}{2}\gamma_{1}R_0^{\frac{9}{2}}+5\gamma_{2}R_0^{4}+\frac{3}{2}\gamma_{3}R_0^{\frac{1}{2}}+\gamma_{4}\cdot
\end{eqnarray}

Replacing Eq.(\ref{fR1}) into Eq.(\ref{rho}) and using
Eq.(\ref{H1}) one can get the EoS parameter of the holographic
$f(R)$-gravity model as
\begin{eqnarray}\label{omega1}
\omega_{R}=-1-\frac{2}{3}\frac{W(R)}{h}
\end{eqnarray}
where
\begin{eqnarray}\label{define}
W(R)= \frac{\frac{11}{2}\zeta R^{\frac{11}{2}}+5\vartheta
R^{5}-\frac{3}{2}\theta^2 R^{3}+\kappa R^{\frac{5}{2}}+\kappa''
R^{2}+ \xi' R^{\frac{3}{2}}+\sigma
R+2(m_{+}\varsigma_{+}+m_{-}\varsigma_{-})+\varepsilon} {\zeta
R^{\frac{11}{2}}+\vartheta R^{5}-\frac{1}{h}\theta^2 R^{3}+\kappa'
R^{\frac{5}{2}}+\kappa'''R^{2}-\xi
R^{\frac{3}{2}}+\sigma'R+\frac{2}{h}(\varsigma_{+}+\varsigma_{-})},
\,\,\,\,\,\,\
\end{eqnarray}
and
\begin{eqnarray}
\zeta &=&\gamma_1 (7h-90),\,\,\,\,\,\,\,\,\,\,\,\ \vartheta
=6\gamma_2
(h-12),\,\,\,\,\,\,\,\,\,\,\,\,\,\,\,\,\,\,\,\,\,\,\,\,\,\,\,\
\theta = \frac{2h}{3(2h+1)},\,\,\,\,\,\,\,\,\,\
\kappa = \sqrt{\frac{h}{6(2h+1)}},\,\,\,\,\ \nonumber\\
\kappa'&=& \frac{4(h-1)}{h(2h+1)}\kappa,\,\,\,\,\,\,\,\,\,\
\kappa'' =-\frac{38h^2+31h+36}{12(2h+1)},\,\,\,\,\,\
\kappa'''=\frac{(h+3)}{h^2}\theta,\,\,\,\,\,\,\,\,\,\,\,\,\,\
\xi =\gamma_3(h+2),\,\,\,\,\ \nonumber\\
\xi'&=& (12h\kappa-\frac{3}{2}\xi),\,\,\,\,\,\,\,\,\,\ \sigma
=\frac{1}{2}(45h-4\gamma_4-9),\,\,\,\,\,\,\,\,\,\,\ \sigma'=
-2h(\gamma_4+8),\,\,\,\,\,\
\varepsilon =-27(2h+1)h,\,\,\,\,\ \nonumber\\
\varsigma_{\pm} &=&((m_{\pm}-2)h-2m_{\pm} ^2
+3m_{\pm}-1)\lambda_{\pm}R^{m_{\pm}}\cdot
\end{eqnarray}

EoS parameter (\ref{omega1}) corresponds to a phantom accelerating
universe, i.e. $\omega_R <-1$ when $\frac{W(R)}{h}>0$. Recent
observational data indicates the EoS parameter $\omega_R$ at the
present lies in a narrow strip around  $\omega_R=-1$ and is quite
consistent with being below this value. So, with
$\frac{W(R)}{h}=0$ and using boundary conditions (\ref{boundary})
one can finds $h=0$ or $h=-\frac{1}{2}$ that is consistent with
EoS parameter of cosmological constant.

For the second class of scale factor, Eq.(\ref{a2}), and using
Eq.(\ref{H2}), the future event horizon $R_h$ reduces to
\begin{eqnarray}\label{Rh}
R_{h}=\frac{1}{h-1}\sqrt{\frac{6h(2h-1)}{R}},\,\,\,\,\,\,h > 1
\cdot
\end{eqnarray}
where the condition $h > 1$ is obtained due to have a finite
future event horizon. If we repeat the above calculations, the
both of $f(R)$ and $\omega_R$ corresponding to the HDE for the
second class of scale factor (\ref{a2}) will been yielded.
Replacing Eq.(\ref{Rh}) into Eq.(\ref{rhoLamda}) yields
\begin{eqnarray}\label{rhoLamda2}
\rho_{\Lambda}=\frac{2\alpha c^2(h-1)^2}{h(2h-1)}R \cdot
\end{eqnarray}

The result for $f(R)$ is same as Eq.(\ref{fR1}) where now
\begin{eqnarray}\label{m2}
m_{\pm}=\frac{3-h\pm\sqrt{1+10h+h^2}}{4},
\end{eqnarray}
and
\begin{eqnarray}\label{gamma2}
\gamma_{1}&=&8\gamma_c h^2\sqrt{\frac{6h}{2h-1}}\frac{\left(m_+-m_-\right)}{(11-2m_+)(11-2m_-)},\nonumber\\
\gamma_{2}&=&\eta\gamma_c\frac{(m_+-m_-)}{(5-m_+)(5-m_-)},\nonumber\\
\gamma_{3}&=&-12\gamma_c (2h-1)\sqrt{\frac{6h}{2h-1}}\frac{\left(m_+-m_-\right)}{(3-2m_+)(3-2m_-)},\nonumber\\
\gamma_{4}&=&18\gamma_c (2h-1)\frac{(m_+-m_-)}{(1-m_+)(1-m_-)},\nonumber\\
\eta &=&(6 c^2+4)h^4-(15 c^2+4)h^3+(12 c^2-1)h^2-3 c^2 h ,\nonumber\\
\gamma_c&=&\frac{1}{3h^2\sqrt{1+10h+h^2}(2h-1)}\cdot
\end{eqnarray}

Also the EoS parameter is obtained as
\begin{eqnarray}\label{omega1}
\omega_{R}=-1+\frac{2}{3}\frac{W(R)}{h},
\end{eqnarray}
but
\begin{eqnarray}\label{define}
\zeta &=&\gamma_1 (7h+90),\,\,\,\,\,\,\,\,\,\,\,\ \vartheta
=6\gamma_2
(h+12),\,\,\,\,\,\,\,\,\,\,\,\,\,\,\,\,\,\,\,\,\,\,\,\,\,\,\,\
\theta = \frac{2h}{3(2h-1)},\,\,\,\,\,\,\,\,\,\
\kappa = \sqrt{\frac{h}{6(2h-1)}},\,\,\,\,\ \nonumber\\
\kappa'&=& \frac{4(h-1)}{h(2h-1)}\kappa,\,\,\,\,\,\,\,\,\,\
\kappa'' =-\frac{38h^2-31h+36}{12(2h-1)},\,\,\,\,\,\
\kappa'''=\frac{(h-3)}{h^2}\theta,\,\,\,\,\,\,\,\,\,\,\,\,\,\
\xi =\gamma_3(h-2),\,\,\,\,\ \nonumber\\
\xi'&=& (12h\kappa-\frac{3}{2}\xi),\,\,\,\,\,\,\,\,\,\ \sigma
=-\frac{1}{2}(45h+4\gamma_4+9),\,\,\,\,\,\,\ \sigma'=
-2h(\gamma_4+8),\,\,\,\,\,\
\varepsilon =27(2h-1)h,\,\,\,\,\ \nonumber\\
\varsigma_{\pm} &=&((m_{\pm}-2)h+2m_{\pm} ^2
-3m_{\pm}+1)\lambda_{\pm}R^{m_{\pm}}\cdot
\end{eqnarray}
which describes an accelerating universe with the quintessence EoS
parameter when $0<\frac{W(R)}{h}<1$, i.e.
$-1<\omega_R<-\frac{1}{3}$

\section{Concluding Remarks}

In this paper, the evolution of the gravitational fields both
analytically and numerically in the $f(R)$ modified gravity model
coupled by the first gravitational theory was considered where it
was invariant under the scale transformation and was presented by
Weyl. A formulation of gravity as a simple modified model
characterized by one scalar field $\phi$ which can be viewed in
our example was taken into consideration as well. Analytical study
of the solution indicates that under special condition, the
universe may undergo a transition from quintessence to phantom
phase which is also supported by numerical analysis. In analytic
studying of the dynamics of the EoS parameter we obtain the
constraints that one has to impose on the scalar field and their
first and second derivatives in order to have phantom crossing. In
numerical approach, the EoS parameter crosses $\omega=-1$ for $t >
0$. We investigated about a bouncing non-singular cosmology, with
an initial contracting phase which lasts until to a non-vanishing
minimal radius is reached and then transits into an expanding
phase which provides a possible solution to the singularity
problem of Standard Big Bang cosmology, a problem which is not
cured by scalar field driven inflationary models. The evolution of
EoS parameter, hubble parameter and scale factor numerically
obtained. The violations of the null energy condition required to
get a bounce are obtained for the model, which allows a transition
of the EoS parameter through the cosmological constant boundary.
The result is that in the analytical discussion of the phantom
crossing behavior of the EoS parameter, we have to also constrain
the scalar field and their first and second derivatives. Besides,
we have also additional constraints on hubble parameter and its
first and second derivatives.

Furthermore, the HDE model which is begun from some important
characteristics of quantum gravity and is motivated from the
holographic hypothesis, was investigated in the framework of
$f(R)$-gravity. A natural unification of the early-time inflation
and late-time acceleration because of different role of
gravitational terms relevant at small and at large curvature and
may naturally describe the transition from deceleration to
acceleration in the cosmological dynamics was given by modified
gravity. The modified gravity based on the $f(R)$, coupled by Weyl
tensor, action in the spatially flat FRW universe for the two
class of scale factor, according to the original version of the
HDE scenario was reconstructed and the EoS parameters of the
corresponding our model  was given. Our considerations indicated,
for the first class of scale factor, the EoS parameter always
crosses the phantom-divide line, whereas for the second class,
behaves like the quintessence. Furthermore our model corresponding
to the HDE can predict the early-time inflation of the universe.

\end{document}